\documentclass[10pt,a4paper,twoside]{article}
\usepackage{epsfig}
\usepackage{baltlat5}
\usepackage{wrapfig}
\begin{document}
\ \
\vspace{-0.5mm}

\setcounter{page}{79}
\vspace{-2mm}

\titlehead{Baltic Astronomy, vol.\ts 17, 79--86, 2008}

\titleb{THE CONTACT BINARY GSC\,04778--00152 WITH\\ 
A VISUAL COMPANION}

\begin{authorl}
\authorb{T.~Tuvikene}{1}
\authorb{T.~Eenm\"ae}{2}
\authorb{C.~Sterken}{1} and
\authorb{E.~Brogt}{3}
\end{authorl}

\begin{addressl}
\addressb{1}{Vrije Universiteit Brussel, Pleinlaan 2, B-1050 Brussel, 
Belgium}

\addressb{2}{Tartu Observatory, 61602 T\~oravere, Estonia}

\addressb{3}{Steward Observatory, Department of Astronomy, 
The University of Arizona, Tucson, AZ 85721, U.S.A.}
\end{addressl}

\submitb{Received 2008 January 10; revised and accepted 2008 March 28}

\begin{summary}
Photometric and spectroscopic observations of the unstudied 12th-magnitude
eclipsing binary GSC\,04778--00152 are presented. We report the discovery of
a visual companion about 1~mag fainter and 2~arcsec away from the
binary. By subtracting the light contribution of the visual companion, we
obtain the $U\!BV\!RI$ light curves of the binary system alone. The shape of
the light curve indicates that GSC\,04778--00152 is an A-type W\,UMa contact 
binary. From light-curve modeling, we derive parameters of the binary system.
\end{summary}


\resthead{The contact binary GSC\,04778--00152}{T.~Tuvikene,
T.~Eenm\"ae, C.~Sterken, E.~Brogt}

\sectionb{1}{INTRODUCTION}
\vspace{-4mm}

The variable star GSC\,04778--00152 ($\alpha_{\rm J2000}=05^{\rm h}31^{\rm m}21^{\rm s}$,
$\delta_{\rm J2000}=-7\degr23\arcmin42\arcsec$) was discovered by the All Sky
Automated Survey (ASAS) and classified as a contact or semi-detached binary with
the period of 0.51746 days (Pojma\'nski 2002). The reported $V$-magnitude
at maximum brightness was 11.95 and the amplitude of variation in $V$ was
0.29~mag.

We detected the variability of GSC\,04778--00152 from wide-field images
of the $\delta$\,Sct star V1162 Ori in January of 2006. Later we conducted
follow-up observations, collecting photometric data in multiple passbands
and a few spectra. All these data will be published in a forthcoming data 
paper (Tuvikene et al., in prep.).

\vspace{2mm}
\sectionb{2}{OBSERVATIONS AND DATA REDUCTION}
\vspace{-4mm}

\subsectionb{2.1}{CCD photometry}
\vspace{-4mm}

CCD photometry was carried out on 31 nights during 5 runs between
2006 January and 2007 October.
We used the 0.41-m Meade telescope at Observatorio Cerro Armazones (OCA),
Chile, the 2.4-m Hiltner telescope at the MDM Observatory, Arizona, USA,
and the 1.0-m telescope at SAAO, Sutherland, South Africa. A Johnson $V$
filter was used at OCA and MDM, while at SAAO the observations were made
with $U\!BV\!RI$ filters.

\begin{wrapfigure}[24]{i}[0pt]{77mm}
\psfig{figure=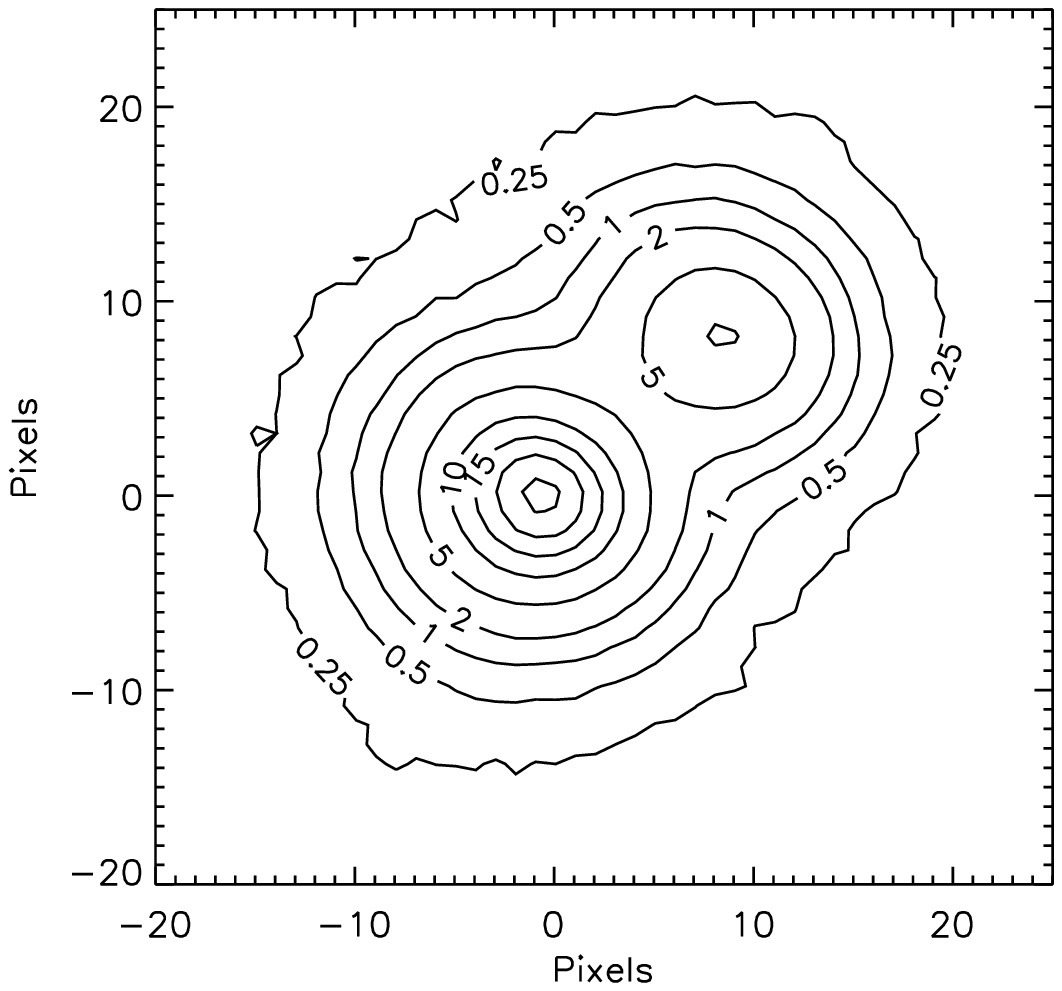,width=75mm,clip=,trim=0 0 0 15}
\vspace{2mm}
\captionb{1}{$V$-band CCD image of GSC\,04778--00152: the binary (star A, 
lower-left) and its visual companion (star B, upper-right). North is up 
and east is left. The contours correspond to constant flux 
(values in 1000$\times$ADU). Image scale is 0.173 arcsec/px.}
\end{wrapfigure}

The CCD images from the MDM Observatory revealed a slightly fainter visual 
companion (hereafter star B) about 2~arcsec away from the eclipsing binary 
(star A). The stars A and B are indicated in Fig.~1.

CCD frames were reduced and aperture photometry was performed as described in 
Tuvikene \& Sterken (2006). Aperture photometry was used to extract the 
magnitudes of the isolated stars in the field, and to measure the total 
brightness of the stars A and B together. In order to extract the magnitudes 
of stars A and B separately, we analysed a subset of CCD frames with PSF 
photometry, using \mbox{\sc daophot} routines in \mbox{\sc eso-midas}.

On the basis of four nights over a 90-day time interval, the visual companion 
B did not show any variability. This allowed us to subtract the contribution 
of star B from the composite light curves acquired with aperture photometry.

The Landolt standard star SA\,98\,193 and about 15 surrounding secondary 
standards were observed in a photometric night. These observations, 
together with the published data from Galad\'i-Enr\'iquez et al.~(2000), were 
used to establish the transformation of the magnitudes to the standard 
$U\!BV\!RI$ system. The standard magnitudes and colours of stars in the 
GSC\,04778--00152 field are presented in Table~1.

\begin{center}
\vbox{\scriptsize
\begin{tabular}{lccccc}
\multicolumn{6}{c}{\parbox{120mm}{
{\normbf \ \ Table 1.}{\norm\ $U\!BV\!RI$ photometry for stars in the 
GSC\,04778--00152 field.}}}\\[2mm]
\tablerule
Star			& $V$		& $B-V$		& $U-B$		& $V-R$		& $V-I$\\
\tablerule
A (min II)		& 12.524~(0.009) & 0.504~(0.010) & 0.088~(0.012) & 0.298~(0.011) & 0.570~(0.011)\\
A (min II) $+$ B	& 12.078~(0.009) & 0.584~(0.010) & 0.140~(0.010) & 0.326~(0.011) & 0.647~(0.011)\\
B			& 13.264~(0.013) & 0.757~(0.017) & 0.271~(0.036) & 0.387~(0.021) & 0.792~(0.017)\\
GSC\,04778--00131	& 12.501~(0.009) & 0.549~(0.010) & 0.052~(0.008) & 0.334~(0.011) & 0.646~(0.011)\\
GSC\,04778--00064	& 13.322~(0.010) & 0.642~(0.011) & 0.101~(0.007) & 0.385~(0.011) & 0.741~(0.012)\\
GSC\,04778--00105	& 13.657~(0.012) & 0.687~(0.013) & 0.095~(0.015) & 0.422~(0.017) & 0.815~(0.016)\\
\tablerule
\end{tabular}
}
\end{center}

Our $V$-band photometry covers 4 primary minima which were used for period
analysis. The times of minimum were determined with the Kwee--van Woerden method 
(Kwee \& van Woerden 1956) and are listed in Table~2.

\begin{center}
\vbox{\small
\begin{tabular}{cccrr}
\multicolumn{5}{c}{\parbox{80mm}{
{\normbf \ \ Table 2.}{\norm\ Heliocentric Julian dates of minima.}}}\\[2mm]
\tablerule
HJD		& Error		& Minimum	& $E$	& $O-C$ \\
		& [days]	&		&	& [days] \\
\tablerule
2453746.61205	& 0.00022	& primary	& 0	&   0.00006  \\
2454080.88796	& 0.00014	& primary 	& 646	& $-0.00015$ \\
2454139.36075	& 0.00020	& primary	& 759	&   0.00020  \\
2454167.30315	& 0.00010	& primary	& 813	&   0.00001  \\
\tablerule
\end{tabular}
}
\end{center}

Using a linear weighted least-squares fit, we obtain the following ephemeris 
for the primary minima:

\begin{equation}
\begin{tabular}{@{$\;$}r @{$\;$}c @{$\;$}r @{$\;$}l}
HJD\,(Min\,I) = 2453746.61199 & + & 0.5174553 & $E$ \\
                 $\pm$0.00021 &   & $\pm$0.0000003 \\
\end{tabular}
\end{equation}

The constant star GSC\,04778--00131 of similar brightness and 
colours as star~A, was used as the comparison star for the binary.
The differential $U\!BV\!RI$ light curves, phased with the
derived period, are plotted in Fig.~2.
The shape of the light curve is typical for W\,UMa type systems.
It can be seen that the eclipses have unequal depths and that 
the secondary minimum displays an interval of constant brightness, 
indicating total occultation. Star~A is therefore an A-type W\,UMa star
where the more massive component has the greater surface brightness.

\begin{figure}
\centerline{\psfig{figure=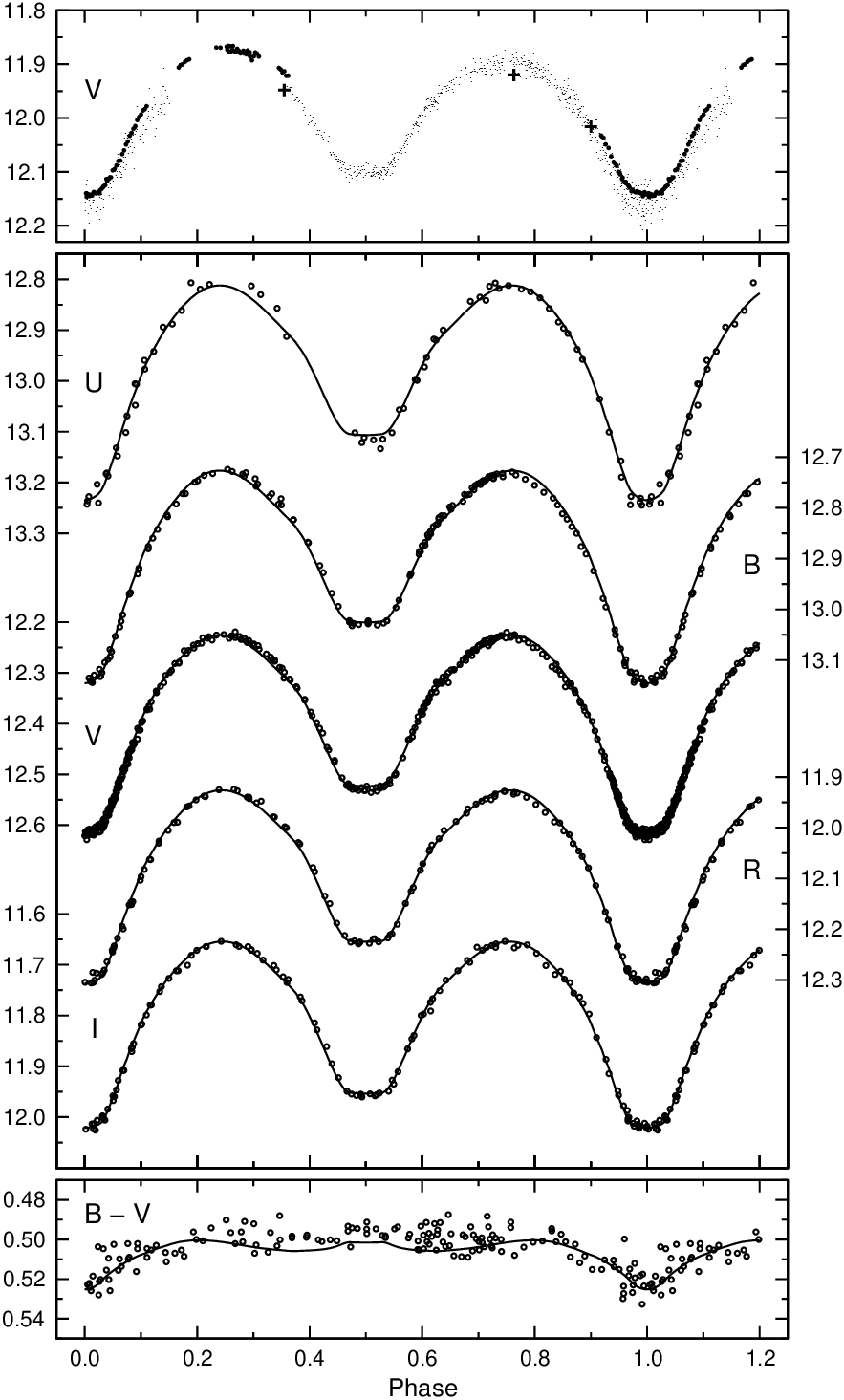,height=165truemm,angle=0,clip=}}
\captionb{2}{Light curves of GSC\,04778--00152. {\sl Top:} differential
light curve including the contribution of star B. Dots refer to the OCA data, 
filled circles to the MDM data and plus signs denote the SAT $y$ magnitudes 
transformed to the Johnson $V$ scale.
{\sl Middle:} phase diagram of the binary star A, with the light of 
star B subtracted (SAAO data). The light curve solution is shown as a
continuous line.
{\sl Bottom:} $B-V$ colour curve of star A (circles), together with the 
synthetic curve (solid line).}
\end{figure}

\vspace{2mm}
\subsectionb{2.2}{CCD astrometry}
\vspace{-4mm}

Astrometric measurements of stars A and B were performed on the basis
of CCD frames from the MDM Observatory. The seeing in these frames varied 
between 1.4 and 1.8~arcsec. The stellar images were fitted with 
PSF, using \mbox{\sc daophot} software. The PSF was constructed from the 
Moffat function on which empirical corrections were applied.

In order to determine the pixel scale and the orientation of the detector, 
we observed 13 astrometric standard stars in the open cluster
NGC\,1647. The coordinates and proper motions of the stars were taken from 
Geffert et al. (1996).

The measurements of star~B relative to star~A are presented in 
Table~3. The columns list the epoch of the measurement, 
the number of CCD frames ($N$), the angular separation with 
standard error ($\rho$, $\sigma_\rho$), and the position angle, measured 
from north to east, with standard error ($\theta$, $\sigma_\theta$). 
The position of star~B in 2007 October coincides, within the errors, with 
the position measured in 2006 December.

\begin{center}
\vbox{\small
\begin{tabular}{@{}lrcccc}
\multicolumn{6}{c}{\parbox{65mm}{
{\normbf \ \ Table 3.}{\norm\ Astrometric measurements of star~B relative to star~A.}}}\\[5mm]
\tablerule
Epoch		& $N$	& $\rho$	& $\sigma_\rho$	& $\theta$	& $\sigma_\theta$ \\
(Julian year)	&	& [$\arcsec$]	& [$\arcsec$]	& [$\degr$]	& [$\degr$] \\
\tablerule
2006.9429	& 21	& 2.131		& 0.002		& 311.67	& 0.08 \\
2007.7947	& 11	& 2.133		& 0.002		& 311.76	& 0.10 \\
\tablerule
\end{tabular}
}
\end{center}

\newpage
\subsectionb{2.3}{Photoelectric photometry}
\vspace{-4mm}

The system was observed with the Str\"omgren Automatic Telescope (SAT) at ESO, 
La Silla at three occasions in 2006 November and 2007 January. A diaphragm of 
17 arcsec was used. Extinction corrections were derived from standard-star 
observations, and transformations to the standard $uvby$ system were based on 
the methods described by Olsen (1994).
The averaged indices at phases 0.35 and 0.70 are: 
$V = 11.93\pm0.01, b-y = 0.377\pm0.009, m_1 = 0.055\pm0.008, c_1 = 0.534\pm0.023$.
Figure~2 (top) shows the 
$y$ magnitudes transformed to the Johnson $V$ scale. There is a systematic 
difference of $0\fm02\pm 0\fm006$ between the photoelectrically determined $V$ 
and the result from CCD imaging. We attribute this difference to the selection
of standard stars, and to different detector and filter characteristics.

\vspace{2mm}
\subsectionb{2.4}{Spectroscopy}
\vspace{-4mm}

Spectroscopic observations were carried out at the Tartu Observatory, Estonia, 
using the 1.5-m telescope with the Cassegrain spectrograph ASP-32 and Andor 
Newton CCD camera. The spectrograph was used in low and moderate resolution 
modes with 600 and 1200~lines/mm gratings. Three regions of the 
spectrum were observed: blue (3635--5785\,\AA), yellow-red (5150--7250\,\AA) 
and near-infrared (7340--9240\,\AA). The spectral resolution varied between
2025 and 4340 and the mean dispersion was between 0.53 and 
1.35\,\AA/px.

Three spectra in the yellow-red and one in the near-infrared region were obtained
in 2007 January and February. The integration time was 50~min for all spectra.
The blue region was covered in 2007 October and November, with 
integration times of 60 and 90~min. The resulting signal-to-noise ratio was 
between 50 and 80, depending on the wavelength.

Due to the average seeing of 
4~arcsec and the 2~arcsec wide slit, which was oriented
in the north-south direction, we also recorded a contribution of 
star~B in the spectra. Being redder than star A (Table~1), 
star~B affects the spectrum in the red region, while in the blue region the 
contamination is not noticeable.

The low-resolution spectrum of the blue region, obtained on 
2007 October~31~/ November~1, was used to derive the $\beta$~index of 
GSC\,04778--00152. The normalised spectrum was convolved with the response 
curves of the H$\beta$ wide and narrow filters of the SAT photometer. 
The instrumental $\beta$~index was then transformed to the standard system, 
using observations of 7 standard stars from Olsen (1983), yielding 
$\beta = 2.658\pm 0.025$.

\begin{figure}
\centerline{\psfig{figure=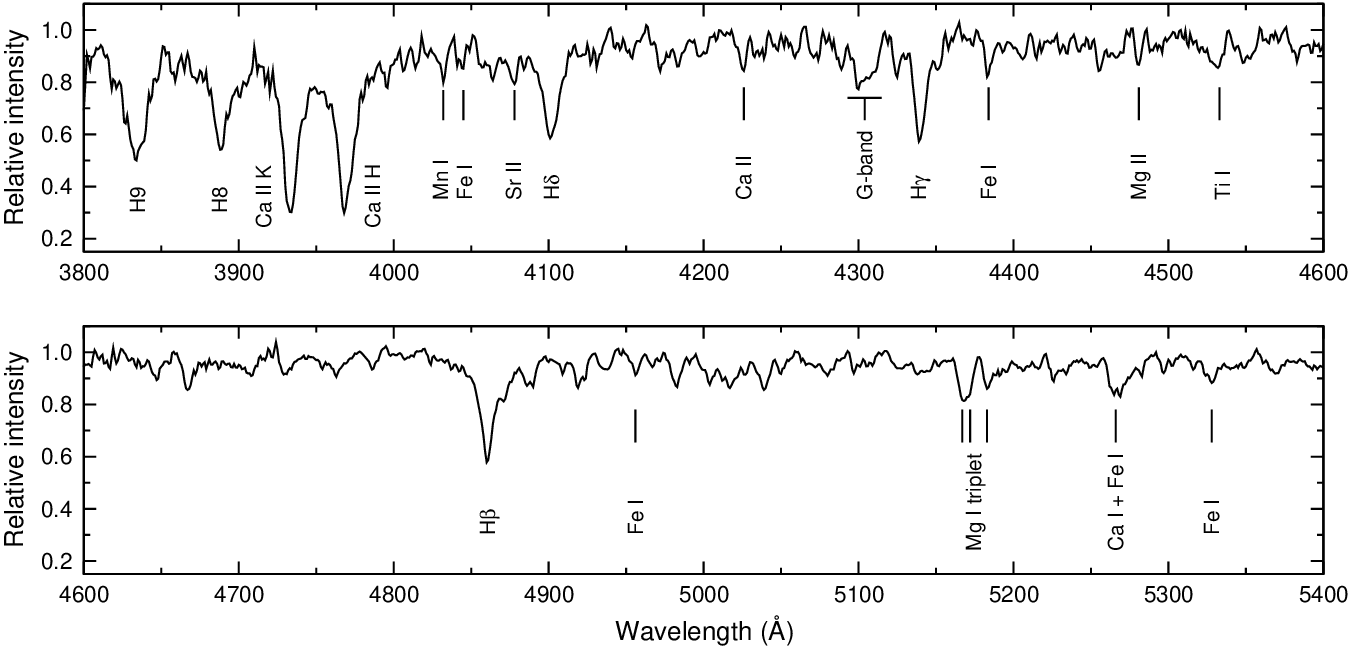,width=120truemm,angle=0,clip=}}
\captionb{3}{The spectrum of GSC\,04778--00152 obtained on 2007 October~31~/ 
November~1, at phase 0.25.}
\end{figure}

\vspace{2mm}
\sectionb{3}{ANALYSIS}
\vspace{-4mm}

\subsectionb{3.1}{Spectral type}
\vspace{-4mm}

From the photoelectric photometry of GSC\,04778--00152 (including star B)
we derived the reddening-free colour indices $[c_1]$ and $[m_1]$ 
(Str\"omgren 1966), which point to a spectral type of F3--F5 for the binary.

Using the Digital Spectral Classification 
Atlas\footnote{\texttt{http://nedwww.ipac.caltech.edu/level5/Gray/Gray\_contents.html}}, 
we analysed the spectrum obtained on 2007 October~31~/ November~1 (Fig.~3).
The strength of the G-band relative to H$\gamma$ and the strengths
of Ca\,{\sc ii} K and H lines lead to a spectral type around F5. The widths 
of the hydrogen lines and their strengths compared to metallic lines also 
refer to F5. Comparing the G-band strength of GSC\,04778--00152 with the 
observed H$\beta$ standards of spectral classes F4V and F6V, places our target 
in between them. There is no evidence that luminosity class is above IV, the 
luminosity-sensitive lines are more similar to those in dwarfs.

We conclude that the spectral class of star~A at 
phase 0.25 is F5V. The uncertainity of our estimation does not exceed one 
subclass. As can be seen from the $B-V$ curve (Fig.~2, bottom), the system 
colour does not change between phase 0.25 and the secondary minimum, where 
we see only the hotter binary component. Therefore the estimated spectral class 
corresponds mainly to the primary star of the binary.

\vspace{2mm}
\subsectionb{3.2}{Stellar parameters}
\vspace{-4mm}

The $uvby$ indices from Section~2.3, together 
with the $\beta$~index from Section~2.4 enable us to determine 
approximate stellar parameters. Applying the Vienna \texttt{TempLogG v2} web 
interface\footnote{\texttt{http://ams.astro.univie.ac.at/templogg/}}, we obtain 
$E_{b-y} = 0.112\pm0.060$, $M_V = 2.7\pm0.5$, $T_{\rm eff} = 6500\pm300$\,K, 
and $\log g = 4.0\pm0.4$. The system is at a distance $d = 560\pm180$\,pc. Note 
that these parameters apply to the composite system A\,+\,B.

The absolute magnitude and distance may also be derived from the luminosity
calibration of contact binaries (Rucinski \& Duerbeck 1997). By using the
intrinsic colour index, which corresponds to the F5V spectral type, 
$(B-V)_0 = 0.44$ (Strai\v zys 1977), and the period value (Section 2.1), we get 
$M_V = 2.7\pm0.4$ and $d = 740_{-110}^{+170}$\,pc.

\vspace{2mm}
\subsectionb{3.3}{Light-curve solution}
\vspace{-4mm}

The $BV\!RI$ light curves of star A were solved simultaneously with the 
binary star modeling package {\sc phoebe}, version 0.29d (Pr\v{s}a \& Zwitter 2005), 
which uses the Wilson-Devinney code as back-end. Curve-dependent weights were 
applied, according to the photometric errors in separate passbands. The 
program was run in the mode ``overcontact binary not in thermal contact''.

Assuming a spectral type of F5V and using the effective 
temperature calibration from Gray et al. (2003), we fixed
$T_{\rm eff,1}=6530\,{\rm K}$ for the primary component. We also adopted the
gravity darkening coefficients, $g_1=g_2=0.32$, and the albedo values, 
$A_1=A_2=0.5$, for the modeling. Subscripts refer to the binary components. 
A logarithmic limb-darkening law was chosen, as suggested by Van Hamme (1993).
The fitted parameters were the mass ratio, $q=m_2/m_1$, the orbital inclination,
$i$, the effective temperature of the secondary, $T_{\rm eff,2}$, the surface 
potential, $\Omega_1=\Omega_2$, and the luminosities of the primary star.
The luminosity in $U$ was fitted separately, using the $U$-band light curve 
alone and fixing all the parameters from the $BV\!RI$ solution.

For error evaluation we carried out a bootstrap-resampling experiment, analogous 
to the one described in Maceroni \& Rucinski (1997). Half of the $BV\!RI$ data 
points were randomly selected and the resampled light curves were 
simultaneously solved with the program {\sc phoebe}, using the complete 
light-curve solution as a starting point. The procedure was repeated 100 times.

The derived binary system parameters are listed in Table~4. We give both sets 
of estimates, the complete light-curve solution with standard errors from
{\sc phoebe}, and the median values of the parameters with uncertainties at
68~percent confidence level from the bootstrap experiment.
The synthetic light and colour curves are plotted in Fig.~2.

\begin{center}
\vbox{\footnotesize
\begin{tabular}{lll@{~~~~~~~~~}lll}
\multicolumn{6}{c}{\parbox{11cm}{
{\normbf \ \ Table 4.}{\norm\ System parameters from the 
light-curve solution of star~A. Subscripts refer to the binary components.}}}\\[5mm]
\tablerule
Parameter	& Solution		& Bootstrap	& Parameter		& Solution	& Bootstrap\\
\tablerule
$q=m_2/m_1$	& 0.1822(7)		& 0.1819(13)	& $\Omega_{\rm in}$	& 2.1878	&\\
$i$ [$\degr$]	& 78.07(17)		& 78.13(27)	& $\Omega_{\rm out}$	& 2.0711	&\\
$T_{\rm eff,1}$ (fixed)	[K]	& 6530	&		& Fill-out		& 0.303(24)	&\\
$T_{\rm eff,2}$ [K]	& 6039(7)	& 6041(10)	& $r_1/a$ (pole)	& 0.5026	& 0.5028(10)\\
$\Omega_1=\Omega_2$	& 2.1524(28)	& 2.1516(50)	& $r_1/a$ (side)	& 0.5516	& 0.5519(14)\\
$[L_1/(L_1+L_2)]_U$	& 0.8725(17)	&		& $r_1/a$ (back)	& 0.5762	& 0.5764(16)\\
$[L_1/(L_1+L_2)]_B$	& 0.8711(14)	& 0.8711(16)	& $r_2/a$ (pole)	& 0.2365	& 0.2365(6)\\
$[L_1/(L_1+L_2)]_V$	& 0.8623(11)	& 0.8623(11)	& $r_2/a$ (side)	& 0.2473	& 0.2474(7)\\
$[L_1/(L_1+L_2)]_R$	& 0.8567(10)	& 0.8567(13)	& $r_2/a$ (back)	& 0.2897	& 0.2900(15)\\
$[L_1/(L_1+L_2)]_I$	& 0.8511(9)	& 0.8511(14)	&			&		&\\
\tablerule
\end{tabular}
}
\end{center}

\vspace{2mm}
\sectionb{4}{CONCLUSIONS}
\vspace{-4mm}

GSC\,04778--00152 is a slightly-reddened eclipsing contact binary with 
a visual companion. Companion B is redder and about 1~mag fainter than binary A.
The stars have an angular separation of 2.132$\pm$0.002~arcsec which, at the 
derived distance of 600--800\,pc, would mean a physical separation of
more than 1000\,AU. It is, therefore, not very likely that stars A and B belong 
to a bound system.

Our spectra indicate that the spectral type of star~A at phase 0.25 is F5V 
which is mainly attributed to the primary component. The corresponding 
effective temperature is 6530\,K. From the light-curve solution we get that 
the secondary component has 18 percent of the mass of the primary and is 
about 500\,K cooler.
\vskip 5mm

ACKNOWLEDGMENTS.
This paper uses observations made at the South African Astronomical 
Observatory (SAAO), Observatorio Cerro Armazones (OCA), the MDM Observatory, 
and the Tartu Observatory.
Part of the work is based on observations made with the Danish 50-cm telescope 
(Str\"omgren Automatic Telescope, SAT) at the European Southern Observatory, 
La Silla, Chile. The telescope is operated by Astronomical Observatory, the 
Niels Bohr Institute, Copenhagen University, Denmark. The authors are indebted 
to Dr. E. H. Olsen for data reduction support. T.~E. would like to thank Dr. 
I.~Kolka for valuable insights into spectroscopy.
This work was supported by the Research Foundation Flanders (FWO), by the 
Flemish Ministry for Foreign Policy, European Affairs, Science, and Technology 
(contract BWS 05--12), and by the Estonian Science Foundation grant No.~6810.

\References

\refb
Galad\'i-Enr\'iquez~D., Trullols~E., Jordi~C. 2000, A\&AS, 146, 169
\refb
Geffert~M., Bonnefond~P., Maintz~G., Guibert~J. 1996, A\&AS, 118, 277
\refb
Gray~R.~O., Corbally~C.~J., Garrison~R.~F., McFadden~M.~T., 
Robinson~P.~E. 2003, AJ, 126, 2048 
\refb
Kwee~K.~K., van Woerden~H. 1956, Bull. Astron. Inst. Netherlands, 12, 327
\refb
Maceroni~C., Rucinski~S.~M. 1997, PASP, 109, 782
\refb
Pojma\'nski~G. 2002, Acta Astron., 52, 397
\refb
Olsen~E.~H. 1983, A\&AS, 54, 55
\refb
Olsen~E.~H. 1994, A\&AS, 106, 257 
\refb
Pr\v{s}a~A., Zwitter~T. 2005, ApJ, 628, 426
\refb
Rucinski~S.~M., Duerbeck~H.~W. 1997, PASP, 109, 1340
\refb
Strai\v zys~V. 1977, in {\it Multicolor Stellar Photometry}, Vilnius, p.~105
\refb
Str\"omgren~B. 1966, ARA\&A, 4, 433
\refb
Tuvikene~T., Sterken~C. 2006, in {\it Astrophysics of Variable Stars},
eds. C.~Sterken, C.~Aerts, ASP Conf. Ser., 349, 359
\refb
Van~Hamme~W. 1993, AJ, 106, 2096

\end{document}